\documentclass[reprint,amsmath,amssymb,aps]{revtex4-1}
\usepackage{graphicx}
\usepackage{dcolumn}
\usepackage{bm}
\usepackage{natbib}
\usepackage{ragged2e}
\usepackage{float}
\usepackage{subfig}
\usepackage{epstopdf}
\usepackage{epsfig}
\usepackage{csquotes}
\usepackage{array}
\usepackage{booktabs}
\usepackage{xcolor}
\begin{document}
\title{Investigating neutron transfer in the $^{9}$Be~+~$^{197}$Au system}

\author{Malika~Kaushik$^{1}$}
\author{S.K.~Pandit$^{2}$}
\author{V.V.~Parkar$^{2}$}
\author{G.~Gupta$^{3}$}
\author{Swati~Thakur$^{1}$}
\author{V.~Nanal$^{3}$}
\author{H.~Krishnamoorthy$^{4,5}$}
\author{A.~Shrivastava$^{2,4}$}
\author{C.S.~Palshetkar$^{3}$}
\author{K.~Mahata$^{2,4}$}
\author{K.~Ramachandran$^{2}$}
\author{S.~Pal$^{6}$}
\author{R.G.~Pillay$^{1}$}
\author{Pushpendra~P.~ Singh$^{1}$}
\email{pps@iitrpr.ac.in}

\affiliation{$^{1}$Department of Physics, Indian Institute of Technology Ropar, Rupnagar - 140 001, Punjab, India}
\affiliation{$^{2}$Nuclear Physics Division, Bhabha Atomic Research Centre, Mumbai - 400 085, India}
\affiliation{$^{3}$Department of Nuclear and Atomic Physics, Tata Institute of Fundamental Research, Mumbai - 400 005, India}
\affiliation{$^{4}$Homi Bhabha National Institute, Anushaktinagar, Mumbai - 400 094, India}
\affiliation{$^{5}$India-based Neutrino Observatory, Tata Institute of Fundamental Research, Mumbai - 400 005, India}
\affiliation{$^{6}$Pelletron Linac Facility, Tata Institute of Fundamental Research, Mumbai - 400 005, India}

\date{\today}%

\begin{abstract}

In this work $\textit{n}$-transfer and incomplete fusion cross sections  for $^{9}$Be~+~$^{197}$Au system are reported over a wide energy range, E$_{c.m.}$ $\approx$ 29-45 MeV. The experiment was carried out using activation  technique and off-line gamma counting. The transfer process is found to be the  dominant mode as compared to  all other reaction channels. Detailed coupled reaction channel (CRC) calculations have been performed for $\textit{n}$-transfer stripping and pickup cross sections. The measured 1$\textit{n}$-stripping cross sections  are explained with CRC calculations by including the ground state and the 2$^{+}$ resonance state (E~=~3.03~MeV) of $^{8}$Be. The calculations for 1$\textit{n}$-pickup, including only the ground state of $^{10}$Be agree reasonably well with the measured cross sections, while it overpredicts the data at subbarrier energies. For a better insight into the role of projectile structure in the transfer process, a comprehensive analysis of 1$\textit{n}$-stripping reaction has been carried out for various weakly bound projectiles on $^{197}$Au target nucleus. The transfer cross sections scaled with the square of total radius of interacting nuclei show the expected Q-value dependence of 1$\textit{n}$-stripping channel for weakly bound stable projectiles.

\end{abstract}
\pacs{Valid PACS appear here}
\maketitle

\section{\label{sec:level1}Introduction}

The nuclear reactions involving weakly bound projectiles open up many distinctive facets of reaction dynamics and play a crucial role in understanding the interplay between quantum tunneling and the breakup process at energies around the Coulomb barrier~\cite{vj}. The striking characteristics of weakly bound projectiles, e.g., low breakup threshold (\textit{E}$_{\rm BU}$) and underlying cluster structure,  significantly affect nuclear dynamics~\cite{bb,lf1}. The weakly bound projectiles like $^{6,7}$Li and $^{9}$Be can break up into their constituent clusters during the interaction process. 
As a consequence of the weak binding of the projectile, one of the projectile fragments following breakup can get absorbed by the target nucleus, known as breakup fusion~\cite{csni, brf, vvpp, dhlu, rr}. Single step cluster transfer from bound state of the projectile to bound/unbound states of the target is also one of the dominant mechanisms, as shown in recent studies~\cite{kcook, kjc,amm}. These two processes are sometimes also  referred to as incomplete fusion (ICF).\\
Complete fusion (CF) is defined as the amalgamation of the projectile as a whole with the target nucleus. It should be pointed out that CF and breakup followed by sequential capture of all the projectile fragments cannot be distinguished experimentally. In the case of weakly bound projectiles, it was suggested that the fusion suppression at the above barrier energies occurs due to the direct breakup of the projectile, while the breakup following transfer is a dominant mechanism at subbarrier energies \cite{md1}. In several existing studies of  $^{6,7}$Li, $^{9}$Be~\cite{rr,dasgupta,sks}, the fusion suppression was attributed to the breakup of the projectile~\cite{dasgupta}. However, in a recent exclusive measurement, Cook {\it et al.}~\cite{kcook} have shown  that  the suppression of CF is primarily a consequence of strong clustering in weakly bound projectiles rather than the breakup prior to reaching the fusion barrier. These recent results  emphasize the role of transfer reactions with weakly bound projectiles.\\

Investigation of the transfer channel measurements provides insight into low energy reaction dynamics, particularly to understand the role of valence nucleons, the spatial correlations of the valence nucleons, and the pairing properties~\cite{al1}. Relatively larger (1\textit{n} and 2\textit{n}) transfer cross sections have been observed in reactions involving halo projectiles ($^{6,8}$He) \cite{an,al1}. Several investigations with stable weakly bound projectile, $^{9}$Be~\cite{ydf,rr,sk,vvpp}, suggest that the neutron transfer channels dominate over CF at subbarrier energies. The CF excitation functions typically show a steeper fall off compared to the transfer channels~\cite{ydf}. In $^{6}$Li~+~$^{96}$Zr~\cite{sph} reaction, it is reported that the neutron transfer channels contribute significantly to the total reaction cross section, especially at energies around the Coulomb barrier. In $^{6}$He~+~$^{64}$Zn experiment, Pietro~$\textit{et al.,}$~\cite{adp1} observed that  transfer and breakup channels comprise the largest fraction of the reaction cross section (about 80$\%$) in near barrier region.\\ 

In a recent measurement of $^{9}$Be~+~$^{197}$Au system~\cite{mk}, the fusion cross sections have been found to be suppressed by $\sim39\%$ w.r.t. theoretical calculations (CCFULL) at above barrier energies. The $^{9}$Be, Borromean nucleus,  may break into $\alpha$~+~$\alpha$~+~\textit{n} (\textit{E}$_{\rm BU}$~=~1.57~MeV)~\cite{pp,tad} or $^{8}$Be~+~\textit{n} (\textit{E}$_{\rm BU}$~=~1.66~MeV) or  $^{5}$He~+~$\alpha$ (\textit{E}$_{\rm BU}$~=~2.31~MeV)~\cite{nk1}. The $^{8}$Be and $^{5}$He are unbound with lifetimes of 10$^{-16}$s and 10$^{-21}$s, respectively. In reactions with $^{9}$Be, at energies around the Coulomb barrier, the \textit{n}~+~$^{8}$Be cluster configuration is shown to play a prominent role as compared to the $^{5}$He~+~$\alpha$ configuration~\cite{vvpp,sk}. In this work, we have investigated neutron transfer and incomplete fusion in the $^{9}$Be~+~$^{197}$Au system at energies around the Coulomb barrier. It should be noted that in inclusive measurements, contributions by different mechanisms  breakup fusion and single step cluster transfer can not be distinguished and are referred as ICF in this paper. The experimental methodology and analysis are presented in section II.  The detailed theoretical model calculations and their comparison with the experimental data of transfer channels cross sections in $^{9}$Be~+~$^{197}$Au system are described in section III. The analysis of transfer channels cross sections in $^{9}$Be~+~$^{197}$Au system (present work) and the reactions involving other weakly bound projectiles, $\rm x$~+~$^{197}$Au systems are delineated in Section IV. The summary of the present work is outlined in section V.

\section{\label{sec:level1}Experimental details and data analysis}

The experiment was performed using 30-47 MeV $^{9}$Be beam from the BARC-TIFR Pelletron Linac Facility, Mumbai, India. A detailed description of the experimental setup and data analysis is given in Ref.~\cite{mk}. The self-supporting $^{197}$Au targets foils (thickness $\sim$ 1.3 - 1.7~mg/cm$^{2}$) together with  the aluminum catcher foils (thickness $\sim$ 1.5~mg/cm$^{2}$) to stop recoiling products were used.  Irradiation was done in 1-2 MeV steps with a beam current of 8-15 pnA. In some cases, stacked target-catcher foil assemblies were deployed, and beam energy at the center of the target was calculated using SRIM \cite{tr}. The activity in the irradiated targets was measured using off-line $\gamma$-ray counting method with efficiency calibrated HPGe detectors. The counting geometry, i.e., sample at a distance of 10~cm from the detector face or at the face, was chosen depending on the activity level in the sample. The transfer and ICF reaction products were identified by their characteristic $\gamma$-rays. The identification was independently confirmed by half-life measurements.

\begin{figure}
    \centering
    \includegraphics[trim=1.4cm 2.0cm 1.4cm 2cm,width=8.5cm]{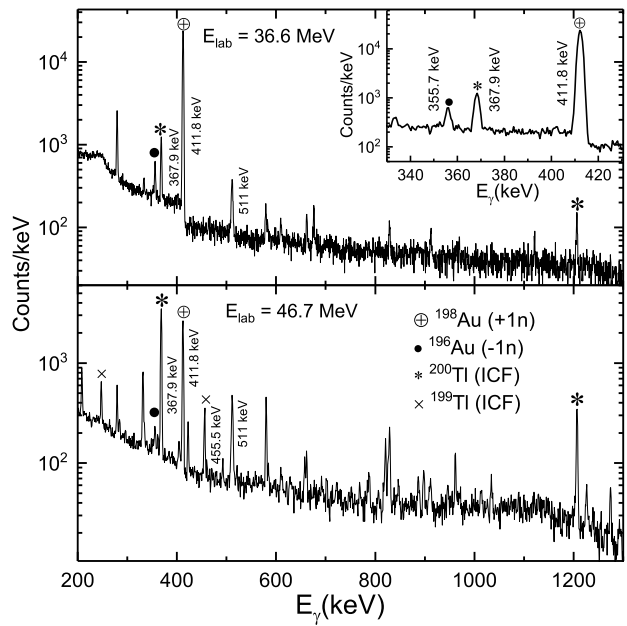}
    \caption{Typical $\gamma$-ray spectra of $^{9}$Be~+~$^{197}$Au system measured at E$_{lab}$~=~36.6 MeV (t$_{collection}$~=~11.1 hr)  and and 46.7 MeV (t$_{collection}$~=~1.55 hr), recorded after a cooldown period 4.64 days and 0.71 days, respectively. Some of intense decay $\gamma$-lines corresponding to different reaction products are marked.}
    \label{fig1}
\end{figure}

Typical $\gamma$-ray spectra  corresponding to  E$_{lab}$~=~46.7 MeV and 36.6 MeV energies are shown in Fig.~\ref{fig1}, where the $\gamma$-rays corresponding to transfer $^{198}$Au (1\textit{n}-stripping), $^{196}$Au (1\textit{n}-pickup), and ICF ($^{199}$Tl, $^{200}$Tl) products are marked. The measured reaction products of transfer and ICF in $^{9}$Be~+~$^{197}$Au system are listed in Table~\ref{table1} along with the most intense characteristic $\gamma$-ray and the corresponding intensity. It may be noted that  $^{199}$Tl can be populated either via $^{197}$Au($^{4}$He,2n) or $^{197}$Au($^{5}$He,3n) reactions. Similarly, both $^{197}$Au($^{4}$He,1n) and $^{197}$Au($^{5}$He,2n) contribute to the production of $^{200}$Tl. \\
The cross sections for the reaction products populated due to transfer and ICF channels are extracted from the photopeak yields of characteristic $\gamma$-rays by taking into account the decay during the irradiation~\cite{mk}. The measured cross sections at different energies are listed in Table~\ref{table3} and plotted in Fig.~\ref{fig2}. It should be mentioned that the quoted  errors include statistical and fitting errors.

\begin{table}[H]
\caption{List of reaction products populated via transfer and ICF channels in $^{9}$Be~+~$^{197}$Au system  with respective half-life (T$_{1/2}$) together with energy and absolute intensity (I$_{\gamma}$) of the most intense  characteristic $\gamma-$ray~\cite{nndc}.}
\begin{ruledtabular}
\begin{tabular}{lllll} 
 Reaction & Nuclide & T$_{1/2}$  & E$_{\gamma}$ (keV) & I$_{\gamma}$($\%$)\\ [0.5ex] 
 \hline
 1n stripping & $^{198}$Au  &  2.69 d  & 411.8    & 95.6 \\

 1n pickup & $^{196}$Au &  6.17 d &  355.7   & 87.0 \\
        
  ICF & $^{200}$Tl  & 26.1 h  &   367.9    & 87.0 \\
 
  ICF & $^{199}$Tl  & 7.42 h  & 455.5    & 12.4    \\
\end{tabular}
\label{table1}
\end{ruledtabular}
\end{table}

\section{\label{sec:level1}Results and discussion}

\begin{figure}[H]
    \centering
    \includegraphics[trim=2cm 2cm 2cm 2.5cm, width=6.5cm]{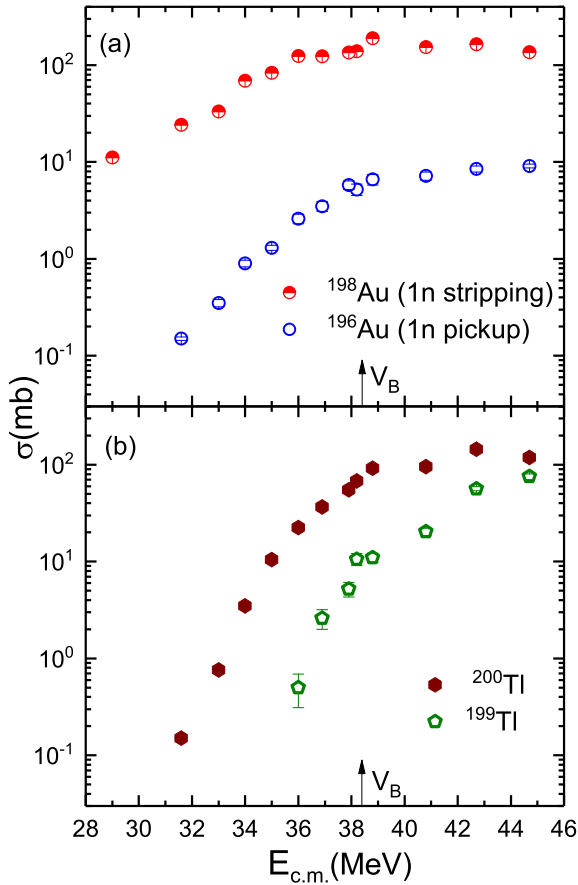}
    \caption{Measured excitation functions of (a) transfer ($^{198,196}$Au), and (b) ICF ($^{200,199}$Tl) channels populated in $^{9}$Be~+~$^{197}$Au system. The barrier value (\textit{V}$_{B}$~=~38.4 MeV) is marked on energy axis, and symbols are self explanatory.}
    \label{fig2}
\end{figure}

\begin{figure}[h]
    \centering
    \includegraphics[trim=3.5cm 2cm 3.5cm 2cm, width=6.5cm]{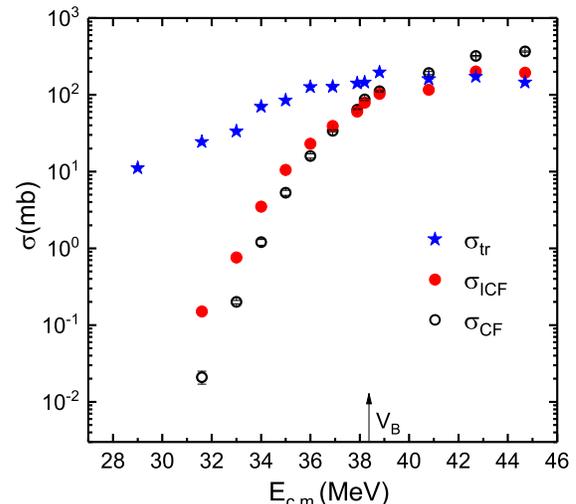}
    \caption{Comparison of measured transfer, ICF, and CF~\cite{mk} cross sections as a function of E$_{c.m.}$ in $^{9}$Be~+~$^{197}$Au system.}
    \label{ICF_TF}
\end{figure}

Excitation functions for reaction products populated due to 1$n$-stripping ($^{198}$Au), 1$n$-pickup ($^{196}$Au) and ICF followed by neutron evaporation ($^{199,200}$Tl) are shown in Fig.~\ref{fig2}(a) and Fig.~\ref{fig2}(b), respectively. The value of barrier (\textit{V}$_{\rm B}$ = 38.4 MeV) \cite{mk} is shown by an arrow. As shown in Fig.~\ref{fig2}, the transfer component is measured down to $\approx$ 29 MeV (i.e., $\approx$ 24 $\%$ below the barrier), while the incomplete fusion component is measured down to $\approx$ 18 $\%$ below the barrier. In order to get a comprehensive picture of CF~\cite{mk}, ICF and transfer reactions,  cross sections of all these channels are plotted as a function of energy in Fig.~\ref{ICF_TF}. As can be seen from this figure, the transfer process significantly dominates over CF and ICF processes at subbarrier energies, whereas CF shows higher cross sections at above barrier energies.

\begin{table*}[t]
\caption{\label{tab:table3} Experimentally measured cross sections of ICF ($^{200,199}$Tl) evaporation residues (ERs) and transfer ($^{198,196}$Au) channels populated in $^{9}$Be~+~$^{197}$Au system, $V_{\rm B}$ = 38.4 MeV.}
\begin{ruledtabular}
\begin{tabular}{cccccc}
  E$_{Lab}$ (MeV) & E$_{c.m.}$ (MeV) & $^{200}$Tl (mb) & $^{199}$Tl (mb) & $^{198}$Au (mb)  & $^{196}$Au (mb)\\ \hline
  
  30.4    & 29.1   & -   & - & 11.2 $\pm$ 0.1 & - \\
  
   33.0    & 31.6 & 0.150 $\pm$0.002 &-   & 24.3 $\pm$ 0.2 &  0.150 $\pm$ 0.006 \\
   
   34.5  & 33.0 & 0.76 $\pm$ 0.01 & -        & 33.2 $\pm$ 0.2 & 0.35 $\pm$ 0.03  \\
   
   35.6  & 34.0 & 3.50 $\pm$ 0.05 & -        &  69.3 $\pm$ 0.4 &  0.90 $\pm$ 0.07 \\ 
   
    36.6  & 35.0 & 10.5 $\pm$ 0.09 & -   &  83.7 $\pm$ 0.3 &  1.30 $\pm$ 0.08 \\
   
    37.6  & 36.0 & 22.5 $\pm$ 0.2 & 0.5 $\pm$ 0.2 &  124.2 $\pm$ 0.6  &   2.6 $\pm$ 0.3\\ 
     
      38.6  & 36.9 & 36.6 $\pm$ 0.5 & 2.6 $\pm$ 0.6  & 123.8 $\pm$ 0.8 &  3.5 $\pm$ 0.4 \\ 
      
       39.6  & 37.9 & 55.2 $\pm$ 0.4 & 5.2 $\pm$ 0.9 & 135.4 $\pm$ 0.6 &   5.8 $\pm$ 0.6 \\ 
      
       39.9  & 38.2 & 68.0 $\pm$ 0.6 & 10.6 $\pm$ 1.5 & 139.7 $\pm$ 1.4 &  5.2 $\pm$ 0.7 \\
      
       40.6  & 38.8 & 92.1 $\pm$ 0.8 & 11.0 $\pm$ 1.2 & 189.6 $\pm$ 1.4 &  6.6 $\pm$ 0.8  \\
    
     42.7  & 40.8 & 95.7 $\pm$ 0.8 & 20.3 $\pm$ 1.9 & 154.1 $\pm$ 1.3 &   7.2 $\pm$ 0.7 \\
     
       44.7  & 42.7 & 144.3 $\pm$ 1.3 & 56.5 $\pm$ 2.8 & 164.7 $\pm$ 1.4  &  8.5 $\pm$ 0.6\\
    
    46.7  & 44.7 & 119.1 $\pm$ 1.2 & 75.4 $\pm$ 5.1 & 136 $\pm$ 1 &  9.1 $\pm$ 0.4 \\ 
\end{tabular}
\label{table3}
\end{ruledtabular}
\end{table*}

\begin{table*}

\caption{\label{table5} Potential parameters used in CRC calculations for $^{9}$Be~+~$^{197}$Au, $^{6,7}$Li~+~$^{197}$Au systems. The radius parameter in the potentials are derived from $R_{i} = r_{i}.A^{1/3}$, where \textit{i~=~R, V, S, C} and \textit{A} is the target mass number.}

\begin{ruledtabular}
\begin{tabular}{lllllllllll}
 System &V$_{R}$ (MeV) &  r$_{R}$ (fm) & a$_{R}$ (fm) & W$_{V}$(MeV) & r$_{V}$(fm) &a$_{V}$(fm) &  W$_{S}$(MeV) &r$_{S}$(fm)  & a$_{S}$(fm) &  r$_{C}$ (fm) \\ \hline
  
 $n$~+~$^{197}$Au~\cite{dgk} & 50.00  \footnotemark[1] & 1.23   & 0.65 & -   &- & - & 6.00  & 1.23 & 0.65  & -\\
    $2n$~+~$^{197}$Au~\cite{dgk} & 50.00 \footnotemark[1] & 1.23   & 0.65 & -   &- & - & 6.00  & 1.23   & 0.65 &  -\\
  $n$~+~$^{196}$Au~\cite{dgk} & 50.00 \footnotemark[1] & 1.23   & 0.65 & -   &- & - & 6.00  & 1.23 & 0.65 &-  \\
  
  $n$~+~$^{8}$Be~\cite{jl} & 50.00 \footnotemark[1] & 1.15   & 0.57 & - & - & - & 5.50  & 1.15 & 0.57  & -\\
  $n$~+~$^{9}$Be~\cite{jl} & 50.00 \footnotemark[1] & 1.15   & 0.57 & - & - & - & 5.50  & 1.15 & 0.57& -  \\

$^{9}$Be~+~$^{197}$Au~\cite{yl} & 257.70    & 1.36   & 0.73 & 16.32   & 1.64 & 0.60 & 46.30  & 1.20 & 0.84  & 1.56\\

    $n$~+~$^{7}$Li~\cite{jc} & 50.00 \footnotemark[1]  & 1.25   & 0.70 & - & - & - & 6.00  & 1.25 & 0.70  & -\\
      $n$~+~$^{6}$Li~\cite{jc} & 50.00  \footnotemark[1] & 1.25   & 0.70 & - & - & - & 6.00  & 1.25 & 0.70  &- \\
      $2n$~+~$^{5}$Li~\cite{jc} & 50.00  \footnotemark[1] & 1.25   & 0.70 & - & - & - & 6.00  & 1.25 & 0.70  & -\\
   $^{6}$Li~+~$^{197}$Au~\cite{jc} & 109.50    & 1.33   & 0.81 & 20.71   & 1.53 & 0.88 &- &- & -& 1.30\\
$^{7}$Li~+~$^{197}$Au~\cite{jc} &  114.20   & 1.29   & 0.85 & 20.71   & 1.74 & 0.81 & -   &- & -& 1.30\\
  
  \footnotetext[1]{Depth adjusted to obtain the correct binding energy.}
 \end{tabular}
\label{table5}
\end{ruledtabular}
\end{table*}

\subsection{\label{sec:level1} CRC calculations: $^{9}$Be~+~$^{197}$Au system}

\begin{figure}[H]
    \centering
    \includegraphics[trim=2cm 2cm 2cm 2.5cm, width=6.5cm]{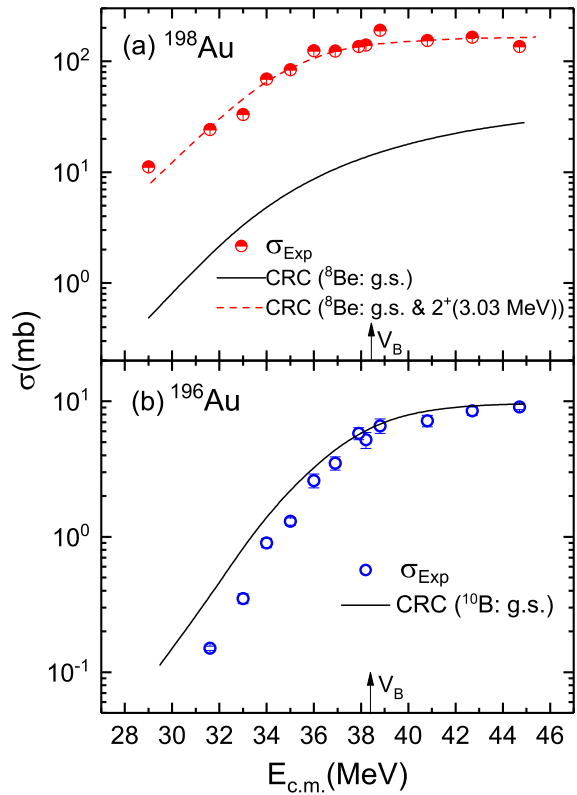}
    \caption{Comparison of measured excitation functions with CRC calculations of transfer channels for (a) $^{198}$Au including the ground state and 2$^{+}$ resonance (E~=~3.03 MeV) state of $^{8}$Be, and only the ground state of $^{8}$Be, and (b) $^{196}$Au with ground state of $^{10}$Be.}
    \label{Fig4}
\end{figure}

For further analysis of transfer channels presented in Fig.~\ref{fig2}, the coupled reaction channel (CRC) calculations have been performed using a theoretical model code FRESCO~\cite{ijt}. For CRC calculations optical model potentials for entrance and exit channels, binding potentials between transferred particle and core nucleus, and spectroscopic factors for different residual states are required inputs. The optical model potential is usually taken to be of Woods-Saxon form for both real and imaginary parts. The potential parameters for $n$~+~$^{197}$Au are taken to be the same as those of $n$~+~$^{208}$Pb from Ref.~\cite{dgk}, since $^{197}$Au and $^{208}$Pb lie in the similar mass region. Since n~+~$^{8}$Be cluster picture of $^{9}$Be has been used in the present analysis, the potential parameters of the binding of $n$ in $^{9}$Be projectile, $\textit{i.e.,}$ the depth of volume potential, radius, diffuseness along with the spin-orbit component are taken from Ref.~\cite{jl}. The global optical model potential parameters for $^{9}$Be~+~$^{197}$Au system are taken from Ref.~\cite{yl}. These potential parameters are further cross-checked by reproducing the experimentally measured elastic scattering cross sections for this system over a wide energy range, E$_{c.m.}$~=~32.5$-$46 MeV~\cite{fg}.\\

The optical model potential parameters for 1\textit{n}-stripping ($^{198}$Au) and 1\textit{n}-pickup ($^{196}$Au) channels calculations are given in Table~\ref{table5}. In 1\textit{n}-stripping, the spectroscopic factor (C$^{2}$S) for $^{9}$Be/$^{8}$Be is taken to be 0.42~\cite{jl} and 1 for the ground state and 2$^{+}$ resonance state of $^{8}$Be, respectively. In 1\textit{n}-pickup, C$^{2}$S for $^{9}$Be/$^{10}$Be is taken to be 1.58~\cite{mbt}. For both 1\textit{n}-stripping and 1\textit{n}-pickup, the spectroscopic factors for target states have been set to 1.0. From the number of $^{198}$Au excited states upto 1.560~MeV given in Ref.~\cite{nndc}, only those states with J value less than 5 and well determined J$^{\pi}$ have been considered. Further, the calculations have been performed with a limited number of states at a time, at three experimental energy points \textit{viz.} the highest (E$_{c.m.}$~=~44.7~MeV), lowest (E$_{c.m.}$~=~33~MeV) and barrier (E$_{c.m.}$~=~38.2~MeV) energy. This procedure has been used to select the states which dominantly contribute to the 1$n$-transfer channel and are listed in Table~\ref{Au_states}. For all the beam energies, the final CRC calculations  have been performed including these states. Similarly, the excited states of $^{196}$Au upto 0.388~MeV having J value less than 5 and well determined J$^{\pi}$ have been considered as listed in Table~\ref{Au_states}. The couplings of $^{198}$Au excited states as given in Table~\ref{Au_states} have been included along with the 0$^{+}$ ground state and 2$^{+}$ resonance state of $^{8}$Be.

\begin{table}
\caption{Energy levels of transfer products used in the CRC calculations in $^{6,7}$Li, $^{9}$Be~+~$^{197}$Au systems~\cite{nndc}.}
\begin{ruledtabular}
\begin{tabular}{ccc rrr}

\multicolumn{2}{c}{$^{199}$Au}& \multicolumn{2}{c}{$^{198}$Au} & \multicolumn{2}{c}{$^{196}$Au}  \\[0.5ex] 
 \hline

 E (MeV) & J$^{\pi}$ & E (MeV) & J$^{\pi}$ & E (MeV) & J$^{\pi}$\\ [0.5ex] 
 \hline
 0.0  &   1.5$^{+}$ & 0.0 & 2$^{-}$  &  0.0  &   2$^{-}$\\

 0.077  & 0.5$^{+}$  & 0.215 & 4$^{-}$ & 0.006  & 1$^{-}$      \\
  0.317  & 2.5$^{+}$&  0.381 & 3$^{+}$  &   0.213  & 4$^{-}$   \\

 0.323 & 1.5$^{+}$  & 0.544 & 4$^{-}$ &  0.234 & 3$^{-}$        \\
0.493 & 3.5$^{+}$&  0.764 & 4$^{-}$  &   0.324 & 1$^{-}$  \\

0.543 & 2.5$^{+}$ & 0.825 & 3$^{+}$ & 0.326    & 1$^{-}$      \\
 
 0.549    & 5.5$^{-}$ & 0.869 & 3$^{-}$  &  0.349    & 2$^{-}$   \\

  0.735 & 3.5$^{-}$& 0.972 & 3$^{-}$ & 0.375 & 3$^{-}$         \\ 
 0.791  & 1.5$^{+}$& 1.032 & 3$^{-}$  & 0.388    & 3$^{+}$   \\

   0.822 &  0.5$^{+}$  & 1.061 & 3$^{-}$ &   &       \\
 
0.907 & 1.5$^{+}$ & 1.115 & 3$^{-}$  &    &   \\

  1.801 &   1.5$^{+}$ & 1.157 & 3$^{-}$ &   &       \\
  2.107  &  0.5$^{+}$ & 1.160 & 3$^{-}$  &    &   \\
 2.205  & 0.5$^{+}$& 1.209 & 3$^{-}$ &   &       \\
 2.412 &   0.5$^{+}$    & 1.240 & 3$^{-}$ &   &       \\
  2.734 & 0.5$^{+}$ & 1.272 & 3$^{-}$  &    &   \\

 & & 1.305 & 3$^{-}$ &   &       \\ 

 & &  1.381 & 3$^{-}$  &    &   \\

  & & 1.396 & 3$^{-}$ &   &       \\
  & &   1.409 & 3$^{-}$  &    &   \\

  & & 1.424 & 3$^{-}$ &   &       \\
 & &  1.444 & 3$^{-}$ &   &       \\
  & &  1.454 & 3$^{-}$  &    &   \\

 & &  1.459 & 3$^{-}$ &   &       \\ 

 & &  1.472 & 3$^{-}$  &    &   \\

 & &  1.496  & 3$^{-}$ &   &       \\
 
  & & 1.560 & 3$^{-}$ &   &       \\

\end{tabular}
\label{Au_states}
\end{ruledtabular}
\end{table}

The excitation functions of (a) 1\textit{n}-stripping ($^{198}$Au), and (b) 1\textit{n}-pickup ($^{196}$Au) channels in $^{9}$Be~+~$^{197}$Au system along with the calculations are shown in Fig.~\ref{Fig4}. As can be seen from the figure, the calculations including couplings only to the ground state of $^{8}$Be underpredict the $\sigma_{\rm 1n-stripping}$ over the entire measured energy range, while those with couplings to both the ground state and 2$^{+}$ resonance state of $^{8}$Be well describe the data. It is evident that the 2$^{+}$ resonance state plays a crucial role in the 1\textit{n}-stripping channel. The Q$_{opt}$~=~0 for $n$-transfer using the semi-classical expression given in Ref.~\cite{wh}. The inclusion of the 2$^{+}$ resonance state of $^{8}$Be at 3.03~MeV provides a better matching  with the Q-value in the 1$n$-stripping reaction.  In the case of $^{196}$Au, the couplings to excited states of $^{196}$Au as given in Table~\ref{Au_states} and 0$^{+}$ ground state of $^{10}$Be have been included in the calculations. As shown in Fig.~\ref{Fig4}(b), the calculations for $^{196}$Au agree reasonably well with the experimental data. However, it overpredicts the measured cross sections at subbarrier energies. It may be noted that $^{198}$Au cross sections are significantly larger than $^{196}$Au cross sections. In principle, $^{198}$Au can also have contribution from the ICF. However, the observed good agreement with CRC calculations indicates that the dominant contribution arises from the transfer process.

\begin{table}
\caption{List of dominant breakup channels together with corresponding breakup threshold energy ($E_{\rm BU}$), ground state Q-values for 1\textit{n}-stripping (Q$_{\rm 1n-strip}$) and 1\textit{n}-pickup (Q$_{\rm 1n-pickup}$) reactions in MeVs for weakly bound projectiles considered in the present study.}
\begin{ruledtabular}
\begin{tabular}{lllll} 
Nuclei & Channel & E$_{\rm BU}$~(MeV) &  Q$_{1n-strip}$ &  Q$_{1n-pickup}$ \\ [0.5ex] 
 \hline
  $^{9}$Be & $\alpha$~+~$\alpha$~+~$n$ & 1.57 & 4.85 & -1.26 \\ 
           & $^{8}$Be~+~$n$  & 1.66 & & \\
 $^{7}$Li &  $\alpha$~+~t  & 2.47 & -0.74  & -6.04\\
  $^{6}$Li & $\alpha$~+~d  & 1.47 & 0.85  & -0.82\\
  $^{10}$B & $^{6}$Li~+~$\alpha$ &  4.46& -1.92  & 3.38\\
           &    $^{9}$B~+~$n$  & 8.44 & &\\
\end{tabular}
\end{ruledtabular}
\label{sep_energy}
\end{table}

\section{\label{sec:level1}Transfer reaction systematics in $\rm x$~+~$\rm ^{197}Au$ systems }

For a comprehensive understanding of projectile structure effect on transfer reactions, it is useful to carry out a systematic comparison for transfer cross sections ($\sigma_{tr}$) of various weakly bound stable projectiles, \textit{i.e.,} $^{6,7}$Li, $^{9}$Be, and $^{10}$B with $^{197}$Au target. For comparison with transfer reactions of other weakly bound stable projectiles, namely, $^{6,7}$Li~\cite{csa}, the experimental excitation functions  have been analyzed  in the framework of CRC calculations following the similar procedure described in the previous section.

\subsection{\label{sec:level1} CRC calculations: $^{6,7}$Li~+~$^{197}$Au systems }

\begin{figure*}
\centering
    \includegraphics[trim=7cm 1cm 7cm 1.5cm, width=5.0cm]{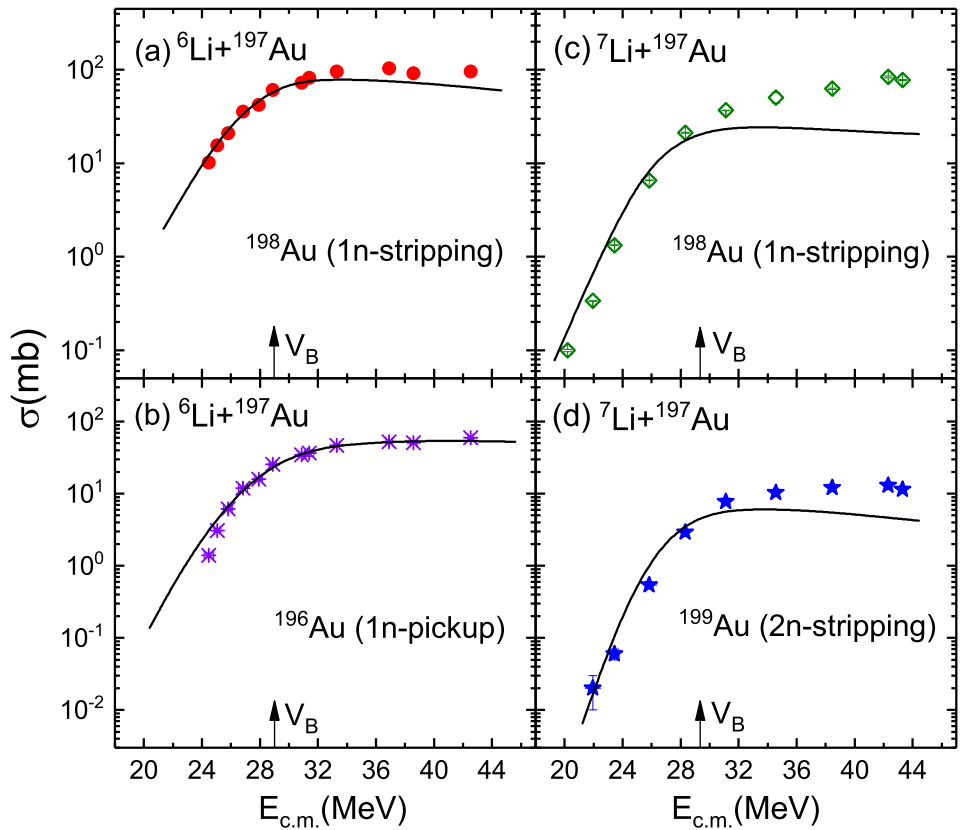}
    \caption {Comparison of measured cross sections~\cite{csa} (a) 1\textit{n}-stripping, (b)1\textit{n}-pickup in $^{6}$Li~+~$^{197}$Au(\textit{V}$_{\rm B}$~=~28.9 MeV) system, and (c) 1\textit{n}-stripping, (d) 2\textit{n}-stripping in $^{7}$Li~+~$^{197}$Au($V_{\rm B}$ = 29.3 MeV) system with CRC calculations (solid curve). Coulomb barrier(\textit{V}$_{\rm B}$) is marked by arrows. }
    \label{fig5}
\end{figure*}

The excitation functions of transfer reaction products, $^{196}$Au, $^{198}$Au  and $^{199}$Au, in $^{6,7}$Li + $^{197}$Au \cite{csa} systems  are presented in Fig.~\ref{fig5}. The global optical potential parameters used in the calculation are presented in Table~ \ref{table5}. The potential parameters: V$_{R}$, W$_{V}$, W$_{S}$ parameter denote the V$_{0}$, W$_{0}$, V$_{S}$ parameters in the $^{6,7}$Li~+~$^{197}$Au systems which are quoted in Ref.~\cite{jc}. In the case of $^{6}$Li~+~$^{197}$Au system, the $^{6}$Li/$^{5}$Li and $^{6}$Li/$^{7}$Li spectroscopic factors are taken to be 1.12 and 1.85 for 1$n$-stripping and 1$n$-pickup channels, respectively \cite{mbt}. Similar to $^{9}$Be~+~$^{197}$Au system, spectroscopic factors for target states ($^{196,198}$Au) have been fixed at 1.0. In 1$n$-stripping calculations, the couplings of $^{198}$Au excited states up to 1.240~MeV as tabulated in Table~\ref{Au_states} and 3/2$^{-}$ ground state of $^{5}$Li have been included.

As shown in Fig.~\ref{fig5}(a), the CRC calculations for 1$n$-stripping ($^{198}$Au) agree well with experimental data at subbarrier energies and underpredict at above barrier energies. Fig.~\ref{fig5}(b) shows the CRC calculations for 1$n$-pickup channel, performed by including couplings of $^{196}$Au excited states up to 0.388~MeV given in Table~\ref{Au_states} and 3/2$^{-}$ ground state of $^{7}$Li. It can be seen that the measured excitation function of the 1$n$-pickup channel is well reproduced by the CRC calculations except for the lowest energy below the barrier.\\
 
A similar analysis has been carried out for $^{7}$Li~+~$^{197}$Au system and presented in Fig.~\ref{fig5}(c)-(d). For CRC calculations of 1\textit{n}-stripping channel ($^{198}$Au), the $^{7}$Li/$^{6}$Li and $^{197}$Au/$^{198}$Au  spectroscopic factors have been taken to be 1.85~\cite{mbt} and 1.0, respectively. The couplings of $^{198}$Au excited states up to 0.972~MeV are listed in Table~\ref{Au_states} and 1$^{+}$ ground state of $^{6}$Li have been included in the CRC calculations. Similarly, for the 2\textit{n}-stripping channel ($^{199}$Au), the couplings of $^{199}$Au excited states up to 2.734~MeV are listed in Table~\ref{Au_states} and 3/2$^{-}$ ground state of $^{5}$Li have been included. All the $^{199}$Au excited states with well determined J$^{\pi}$ values have been considered. The $^{7}$Li/$^{5}$Li and $^{197}$Au/$^{199}$Au spectroscopic factors have been taken as 1.0 in the calculations. As shown in Fig.~\ref{fig5}(c)-(d), the CRC calculations show a reasonable agreement with the experimental data at subbarrier energies and underpredict the data at above barrier energies. The increasing trend of underprediction of CRC calculations at above barrier energies indicate the involvement of more states (high excitation energy) of target like nuclei, which could not be included in the present calculations.\\

\begin{figure}
    \centering
    \includegraphics[trim=2.8cm 2cm 2.5cm 2.5cm, width=5.5cm]{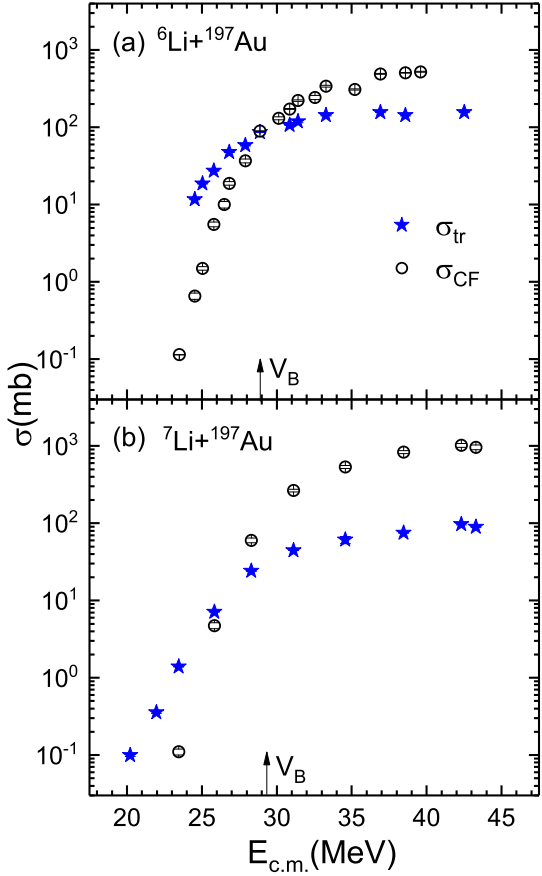}
    \caption{Comparison of the measured excitation functions of CF and transfer reaction  in (a) $^{6}$Li~+~$^{197}$Au, and (b) $^{7}$Li~+~$^{197}$Au systems~\cite{csa}, where \textit{V}$_{\rm B}$ is the barrier. }
    \label{fig6}
\end{figure}

As an illustration of the influence of the projectile structure in transfer reactions, a systematic comparison of $\sigma_{tr}$ with  CF  cross sections ($\sigma_{CF}$) \cite{csa} for sub to above barrier energies has been carried out for $^{6,7}$Li~+~$^{197}$Au systems  and is shown in Fig.~\ref{fig6}.  It can be seen that for both $^{6,7}$Li projectiles  $\sigma_{tr}$ is higher than $\sigma_{CF}$ at subbarrier energies as expected. The $\sigma_{tr}$/$\sigma_{CF}$ is larger for $^{6}$Li than that for $^{7}$Li, but is considerably smaller than that for $^{9}$Be (refer to Fig.~\ref{ICF_TF}). This enhancement of transfer cross section over CF  may be attributed to the structural differences of $^{6,7}$Li, $^{9}$Be projectiles. In $^{9}$Be the last valence neutron has a large spatial extension~\cite{csni} and in addition the coupling to the 2$^{+}$ resonant state in $^{8}$Be plays a crucial role.

\subsection{\label{sec:level1} Comparison of transfer cross sections in $\rm x$~+~$\rm ^{197}Au$ systems}

\begin{figure}
    \centering
    \includegraphics[trim=3cm 1.5cm 3cm 2cm, width=7cm]{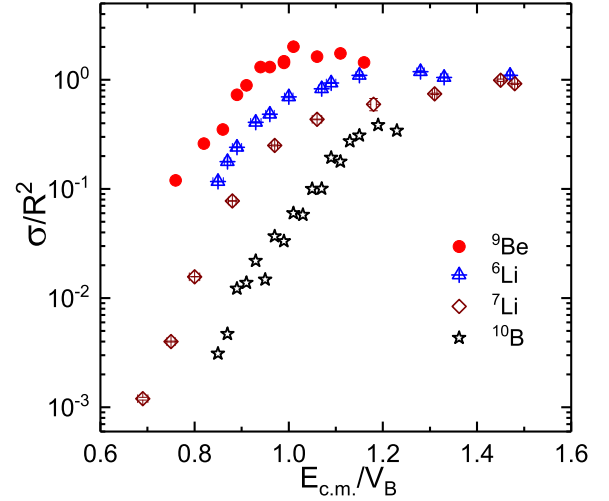}
    \caption{Comparison of measured transfer excitation functions of 1$n$-stripping ($^{198}$Au) channel in $^{9}$Be~+~$^{197}$Au (present work) and other reactions, $\rm x$~+~$^{197}$Au, involving $^{7}$Li~\cite{csa}, $^{6}$Li~\cite{csa}, $^{10}$B~\cite{ma} projectiles.}
    \label{Fig7}
\end{figure}

The excitation functions of the 1\textit{n}-stripping channel with different weakly bound stable projectiles $^{6,7}$Li, $^{9}$Be (present work) and  $^{10}$B on $^{197}$Au target are compared in Fig.~\ref{Fig7} with suitable scaling. In order to suppress trivial differences that arise due to the geometrical effects of the interacting nuclei, a simple semi-classical scaling procedure used in  Ref.~\cite{prsg} for fusion cross section comparison is adopted for the transfer cross section. The scaled parameters are defined as,
\begin{equation}
       \sigma_{sc}~=~\sigma/R^{2}  ~~~{\rm and} ~~~  E_{sc}~=~E_{cm}/V_{B},
\end{equation}
where $R~=~R_{P}~+~R_{T},$ R$_{P}$ is the projectile radius  and R$_{T}$~=~7.27 fm (1.25.A$_{T}^{1/3}$) is the target nucleus radius, and $V_{\rm B}$ is the barrier. The $V
_{\rm B}$ for $^{6,7}$Li, $^{9}$Be~+~$^{197}$Au systems is taken from the  CCFULL calculations, while that for $^{10}$B~+~$^{197}$Au is taken from Ref.~\cite{ma}. The $R_{P}$ values of different projectiles are listed in Table~\ref{radii}. 

\begin{table}
\caption{\label{tab:table5} Radii of different projectiles in $\rm x$~+~$^{197}$Au systems.}
\begin{ruledtabular}
\begin{tabular}{ccc}
Projectile &  R$_{P}$ (fm) & Ref.  \\ \hline
    $^{6}$Li & 2.09     & \cite{tk}  \\
      $^{7}$Li & 2.23   & \cite{tk}  \\
  $^{9}$Be & 2.45    & \cite{tk}  \\
$^{10}$B &  1.95   & \cite{kccp} \\
\end{tabular}
\label{radii}
\end{ruledtabular}
\end{table}

The projectile breakup thresholds ($E_{\rm BU}$) and  1\textit{n}-stripping Q-value are tabulated in Table~\ref{sep_energy}. It can be seen from Fig.~\ref{Fig7} and Table~\ref{sep_energy} that amongst the weakly bound stable projectiles, $\sigma_{sc}$ is the lowest for $^{10}$B, which has the largest negative Q value and highest for $^{9}$Be that corresponds to the largest positive Q value. Further, it should be mentioned that the choice of $R_{P}$ is not crucial in the analysis, as $\sigma$ scaled with $R_{B}^{2}$ (where $R_{B}$ is the barrier radius obtained from CCFULL calculations) also shows a similar trend.

\section{\label{sec:level1}Summary and Conclusions}
In summary, the transfer and ICF reaction cross sections have been measured in $^{9}$Be~+~$^{197}$Au system in an energy range 0.76 $\leq$ E$_{c.m.}$/\textit{V}$_{\rm B}$ $\leq$ 1.16, and analyzed using coupled reaction channel (CRC) calculations. The experimental excitation function of $^{198}$Au populated via 1\textit{n}-stripping channel is well reproduced with the CRC calculations using couplings to the ground state and 2$^{+}$ resonance state of $^{8}$Be. It is shown that the 2$^{+}$ resonance state of $^{8}$Be plays a significant role in 1$n$-stripping channel. For 1\textit{n}-pickup channel, the CRC calculations including only the ground state of $^{10}$Be are in reasonable agreement with the data  at above barrier energies, but overpredict the measured cross sections at subbarrier energies. A systematic analysis of 1$n$-transfer, CF and ICF reaction channels, shows the dominance of 1$n$-transfer over ICF and CF at subbarrier energies in $^{9}$Be~+~$^{197}$Au system.\\

Additionally, the CRC calculations have been performed to analyze the excitation functions of transfer channels in $^{6,7}$Li~+~$^{197}$Au systems~\cite{csa}. In $^{6}$Li~+~$^{197}$Au system, it has been found that the CRC calculations for 1\textit{n}-stripping channel agree well with the experimental data at subbarrier energies but underpredict at above barrier energies, while the 1\textit{n}-pickup channel show excellent agreement over most of the measured energy range. The calculations of 1\textit{n} and 2\textit{n}-stripping transfer channels populated in $^{7}$Li~+~$^{197}$Au system~\cite{csa} agree well with the experimental data at subbarrier energies.\\ 

 In the systematics for transfer reactions, the projectile structure effect is illustrated by comparison of 1n-stripping cross sections of $^{6,7}$Li, and $^{9}$Be projectiles on $^{197}$Au target. The scaled excitation functions of 1\textit{n}-stripping in the $\rm x$~+~$^{197}$Au systems show the highest cross section values for $^{9}$Be projectile,  as expected from the ground state Q-value systematics for the weakly bound stable projectiles.

\section{\label{sec:level1}Acknowledgments}
We thank the PLF staff for providing the steady and smooth beam during the experiments and the target laboratory personnel for their help in the target preparation. We acknowledge the support of the Department of Atomic Energy, Government of India, under Project No. 12-R$\&$DTFR-5.02-0300.

\end{document}